\documentclass[twocolumn,showpacs,preprintnumbers,amsmath,amssymb]{revtex4}

\usepackage{graphicx}
\usepackage{dcolumn}
\usepackage{bm}

\begin{document}


\title{Multiple currents in charged beams}

\author{D.A. Burton}
\author{J. Gratus}
\author{R.W. Tucker}
\affiliation{%
Department of Physics, Lancaster University, UK\\ \& The Cockcroft
Institute, UK
}%

\date{\today}

\begin{abstract}
It is argued that continuum realisations of distributions of
collisionless charged particles should accommodate a dynamically
evolving number of electric currents even if the continuum is
composed of only one species of particle, such as electrons.
A model is proposed that self-consistently
describes the interaction of such a continuum and its electromagnetic
field. It is formulated using a Lagrangian approach and employs a
``folded'' flow map to describe the bulk particle motion.
An asymptotic perturbation
scheme is developed to analyse ultra-relativistic multi-component
current configurations. The model is fully
relativistic and is formulated over Minkowski spacetime using
intrinsic tensor field theory.     
\end{abstract}


\newcommand{\dual}[1]{{\widetilde{#1}}}
\newcommand{\man}{{\cal M}_4}
\newcommand{\Pman}{{\cal M}_2}
\newcommand{\Bman}{{\cal B}_4}
\newcommand{\PBman}{{\cal B}_2}
\newcommand{\UBman}{{\cal B}_3}
\newcommand{\PUBman}{{\cal B}_1}
\newcommand{\VnormJ}{{|J|}}
\newcommand{\dualV}{{\dual V}}
\newcommand{\Real}{{\mathbb{R}}}
\newcommand{\myC}{{\Gamma}}
\newcommand{\UP}{{\underline{P}}}

\newcommand{\calJ}{{\cal J}}
\newcommand{\UcalJ}{{\underline{\cal J}}}
\newcommand{\calF}{{\cal F}}
\newcommand{\calG}{{\cal G}}
\newcommand{\calE}{{\cal E}}
\newcommand{\hatE}{{\hat E}}
\newcommand{\calV}{{\cal V}}
\newcommand{\calW}{{\cal W}}
\newcommand{\calU}{{\cal U}}
\newcommand{\calB}{{\cal B}}
\newcommand{\calL}{{\cal L}}
\newcommand{\calA}{{\cal A}}
\newcommand{\calN}{{\cal N}}

\newcommand{\Tens}{{\otimes}}
\newcommand{\PD}{{\partial}}
\newcommand{\widedual}[1]{{\widetilde{#1}}}
\newcommand{\Vep}{{\varepsilon}}
\newcommand{\zetaVep}{{\zeta^\Vep\left(\hat{\sigma}^\Vep(t,z)\right)}}
\newcommand{\Lie}{{\cal L}}

\newcommand{\Mm}{{\frac{m_0 c^2}{\epsilon_0}}}

\newcommand{\Tensor}{\otimes}
\newcommand{\Pg}{\overline{g}}
\newcommand{\Pnabla}{\overline{\nabla}}
\newcommand{\DivV}{\theta}
\newcommand{\CurveOp}{{\cal R}}

\newcommand{\Parman}{{\cal P}}                           
\newcommand{\Fieldsysrefs}{(\ref{np_maxwell_V_d_F}-\ref{np_maxwell_V_V_dot_V})~}
\newcommand{\Fieldsysrefsnotilde}{(\ref{np_maxwell_V_d_F}-\ref{np_maxwell_V_V_dot_V})}
\newcommand{\Multisysrefs}{(\ref{multi_C_normalisation}-\ref{multi_gauss_ampere-maxwell})~}
\newcommand{\Multisysrefsnotilde}{(\ref{multi_C_normalisation}-\ref{multi_gauss_ampere-maxwell})}

\newcommand{\VarC}{{\cal C}}
\newcommand{\VarCdot}[1]{{{\cal C}^{{#1}\prime}}}
\newcommand{\VarCddot}[1]{{{\cal C}^{{#1}\prime\prime}}}
\newcommand{\VarF}{{\cal F}}
\newcommand{\VarDelta}{{\varepsilon \Delta}}
\newcommand{\VarCinds}[1]{{\cal C}_{[#1]}}
\newcommand{\VarCindx}[2]{{{\cal C}_{#1[#2]}}}
\newcommand{\VarCdotind}[2]{{\cal C}^{{#1}\prime}_{#2}}
\newcommand{\VarCddotind}[2]{{\cal C}^{{#1}\prime\prime}_{#2}}
\newcommand{\Brho}{{\varrho}}

\maketitle
\section{Introduction}
\label{ch_intro}
Progress in high-energy physics relies on accelerator designers contemplating
charged particle beams of ever higher energies and intensities.
As schemes for accelerating charged particles become more complex and ambitious in their aims, it
is apparent that some existing theoretical models are inadequate for a
proper understanding of new challenges. In many existing models, matter
is represented using classical point particles and it is
not clear how to unambiguously and consistently model their
electromagnetic interaction.  

The nub of the problem is precisely how one should sensibly
formulate the interaction of a classical point charge with its \emph{own electromagnetic
field}. The charge density of a point particle is singular and
must therefore be handled carefully. Assumptions must be made about
how the singular Coulombic stresses are compensated by
non-electromagnetic stresses when calculating the particle's
self-force. This issue was addressed by
Dirac~\cite{dirac:1938} nearly 70 years ago and led
to the covariant Lorentz-Dirac equation for the trajectory of a point charge in
an external electromagnetic field. However, unlike more familiar
equations of particle mechanics, the Lorentz-Dirac equation
is a \emph{third-order} ordinary
differential equation in proper time for the particle's
trajectory and has a number 
of unusual properties including self-acceleration and
pre-acceleration. Unless special final conditions are adopted, it predicts that an isolated and free
point charge with non-zero initial velocity will accelerate forever
and if the particle is subjected to a sharp electromagnetic pulse it
will begin to move before the pulse reaches it. Methods for evading such
unpalatable conclusions involve iterating the Lorentz-Dirac equation
in powers of its charge $q_0$.
Landau and Lifshitz~\cite{landau_lifshitz:1962} showed
that truncating the resulting series to any \emph{finite} order leads
to a second-order evolution equation yielding particle trajectories
with more reasonable properties. Although their
argument yields a workable scheme, it is not at all clear that it may be
generalised to a collection of accelerating high-energy charged particles
in close proximity. Neglecting higher-order terms
in $q_0$ may be suspect when the particle number density is
sufficiently high. In conclusion, a number of ad-hoc choices must be made
to obtain a sensible relativistic equation of motion for a collection of charged
point particles starting from first principles. For a recent account of the issues concerning the
derivation of the Lorentz-Dirac equation see~\cite{burton_gratus_tucker:2006}.

Many of the above issues are due to the uneasy marriage of field and
point particle concepts in classical electrodynamics. Classical point
particles are a convenient idealisation and one might argue that quantum
theory must be invoked to yield a palatable answer. However, the testy
relationship between fields and point particles resurfaces in quantum
electrodynamics where divergences in ``bare'' quantities must be
regulated prior to renormalisation to obtain physical results.
 
Given the above complexities and reservations, an alternative approach
has recently been developed~\cite{burton_gratus_tucker:2006}
to analyse the ultra-relativistic dynamics of a collection of
accelerating charged particles. The attitude adopted
in~\cite{burton_gratus_tucker:2006} and the present article is that
models of matter based on classical relativistic field
theory are more suitable for high-energy beam dynamics than those
employing classical point particle notions. The model
in~\cite{burton_gratus_tucker:2006} employs a smooth relativistic
field description of the electromagnetic \emph{and} matter content
where the total energy and momentum of the electromagnetic and matter
fields are conserved. Charged matter is modelled as a $4$-vector field
on spacetime whose trajectories describe the bulk particle
motion.

The partial differential equations governing the electromagnetic and matter
fields in this article are fully coupled
and non-linear. Although exact solutions describing highly symmetric
configurations can be found, the system of equations is in general
only tractable when subjected to an approximation scheme. Such a scheme,
based on a covariant asymptotic expansion in a running parameter
$\varepsilon >0$ around the light-cone, was
introduced in~\cite{burton_gratus_tucker:2006} and permits calculation
of the detailed dynamics of ultra-relativistic charged particle
beams in external electromagnetic fields.
In principle, one can calculate the field quantities to any desired
finite order in $\varepsilon$.

The present article focuses on some of the issues encountered when
analysing the model in~\cite{burton_gratus_tucker:2006}.
Specifically,
the non-linearities in the field equations may lead to solutions whose
charge density diverges despite being initially
regular. Sections~\ref{section:cold_fluid_model} and~\ref{section:development_of_singularities_in_rho} briefly review the charged
continuum model in~\cite{burton_gratus_tucker:2006} and show that diverging
solutions exist satisfying a substantial range of initial conditions. 
This behaviour is commonplace in many continuum models in physics, such as in
neutral gas dynamics and fluid dynamics, and is often
ameliorated by including dissipative
processes. Section~\ref{section:discussion_of_crossing_trajectories}
argues that dissipative processes will not prevent the formation of
multiple currents in charged beams.
Sections~\ref{section:lagrangian_cold_charge}
and~\ref{section:multi_component_charge} develop and analyse a
continuum model accommodating currents with a dynamical number of
components and section~\ref{section:ultra-relativistic_approximation}
extends the ultra-relativistic analysis methods introduced
in~\cite{burton_gratus_tucker:2006} to multi-component charged currents.  
\section{Single current charged continua}
\label{section:cold_fluid_model}
The model discussed in this section and in~\cite{burton_gratus_tucker:2006}
describes a collection of accelerating
charged particles, with (rest) mass $m_0$ and charge $q_0$, as a
dynamical continuum. The vector
field $V$ is the $4$-velocity of the continuum on spacetime and its
integral curves describe the bulk motion of the collection of charged
particles. The scalar field $\calN$ is the particle number density measured by a
comoving observer and $\rho=\frac{q_0^2}{\epsilon_0 m_0 c^2}\calN$ is
called the \emph{reduced} proper charge density where $\epsilon_0$ is the permittivity
of the vacuum and $c$ is the speed of light in the vacuum. In what
follows, units are chosen in which $c=1$. 

The antisymmetric rank $2$ covariant tensor field $F$ (a $2$-form) encodes the
electromagnetic field and the triple $(V,\rho,F)$ satisfies the
covariant Maxwell equations~\cite{benn_tucker:1987}
\begin{align}
&d F = 0\,,
\label{np_maxwell_V_d_F}
\\
&d \star F = - \rho \star \dualV \,,
\label{np_maxwell_V_d_star_F}
\end{align}
on Minkowski spacetime $(\man,g)$ where $g$ is the metric tensor. In
an inertial Cartesian coordinate system $(t,x,y,z)$ in the laboratory frame
\begin{equation*}
g = -dt\Tensor dt + dx\Tensor dx + dy\Tensor dy + dz\Tensor dz,
\end{equation*}
$d$ is the exterior derivative,
$\star$ is the Hodge map associated
with $g$ and the $1$-form $\dualV$ is defined by the property
$\dualV(X) = g(V,X)$ for any vector field $X$.
The field equations for $V$ are obtained using
energy-momentum conservation $d\tau_K=0$ where the total
stress-energy-momentum $3$-form $\tau_K$ is the sum of matter and electromagnetic
contributions :
\begin{equation*}
\tau_K = \rho g(V,K)\star\dualV +
\frac{1}{2}\big(i_K F\wedge\star F - F\wedge i_K \star F\big)
\end{equation*}
and the vector field $K$ is a spacetime translation
on $\man$. Setting $K$ to $\PD_t$, $\PD_x$, $\PD_y$ and $\PD_z$ in $d\tau_K=0$
yields the $\PD_t$, $\PD_x$, $\PD_y$ and $\PD_z$ components of the equation 
\begin{equation}
\nabla_V \dual{V} = \; {i_{V} F}\,,
\label{np_maxwell_V_lorentz}
\end{equation}
where
\begin{equation}
g(V, V) = -1
\label{np_maxwell_V_V_dot_V}
\end{equation}
with $\nabla$ the Levi-Civita connection on $\man$ and $i_V$
the interior (contraction) operator on forms.
The term $i_V F$ in (\ref{np_maxwell_V_lorentz}) is a continuum
generalisation of the covariant expression for the Lorentz force on a
point charge (the charge to mass ratio $q_0/m_0$ has been absorbed
into the definitions of $\rho$ and $F$) where the tangent to the point
charge's proper-time parametrised worldline has been replaced by $V$.
The charged matter
drives the electromagnetic field through (\ref{np_maxwell_V_d_star_F})
and the electromagnetic field acts back on the matter through
(\ref{np_maxwell_V_lorentz}) conserving total energy and momentum.

Equations
\Fieldsysrefs are well-known in charged plasma physics and are often said to
describe a ``cold charged fluid''. They have found application to
accelerator physics~\cite{davidson:2001} in recent years and have
proved useful for examining the stability of high intensity particle beams.  
\section{Development of singularities in the charge density}
\label{section:development_of_singularities_in_rho}
In~\cite{burton_gratus_tucker:2006} highly symmetric exact solutions to
\Fieldsysrefs describing ``walls of
charge'' were used to motivate a hierarchy of field equations for
modelling ultra-relativistic charged particle beams.
Exact solutions to \Fieldsysrefs of the form
\begin{align}
\notag
&F = \calE(t,z)\, dt\wedge dz,\\
\label{wall-of-charge_ansatz}
&V = \frac{1}{\sqrt{1-\mu^2(t,z)}}\left(\PD_t + \mu(t,z)\PD_z\right)
\end{align}
were sought, where $\mu\PD_z$ is the Newtonian velocity of the charge distribution
as measured by the laboratory observer $\PD_t$.
Using
(\ref{np_maxwell_V_d_F}-\ref{wall-of-charge_ansatz}) it follows that
\begin{align}
\label{wall-of-charge_dE}
&d\calE = \rho\# \dual{V},\\
\label{wall-of-charge_lorentz}
&\Pnabla_V \dual{V} = \calE \#\dual{V},\\
\label{wall-of-charge_VV}
&\Pg(V,V)=-1
\end{align}
where the projected metric $\Pg$ is
\begin{equation*}
\Pg = - dt \Tensor dt + dz\Tensor dz,
\end{equation*}
$\#$ is the Hodge map associated with the volume $2$-form
$\# 1 \equiv dt\wedge dz$, $\Pnabla$ is the Levi-Civita connection
of $\Pg$ and $\dual{V}(X)=\Pg(V,X)$ for any vector $X$ on
2-dimensional Minkowski spacetime $\Pman$ with metric $\Pg$.
Equation (\ref{wall-of-charge_dE}) implies
\begin{equation}
V\calE = 0
\label{equation_for_E_along_V}
\end{equation}
i.e. the electric field is constant along the integral curves
of $V$ and so, using (\ref{wall-of-charge_lorentz}), the magnitude
of the acceleration $\Pnabla_V V$ is constant along the integral curves of $V$.
Therefore, the continuum undergoes local hyperbolic motion and it
is straightforward to solve 
to (\ref{wall-of-charge_dE}-\ref{wall-of-charge_VV}) in a comoving
coordinate system $(\tau,\sigma)$ adapted to $V$, as shown
in~\cite{burton_gratus_tucker:2006}, where $V=\PD_\tau$ and $z=\sigma$
on the initial hypersurface $\tau=0$. For charge distributions
initially at rest the Jacobian
of the transformation between $(\tau,\sigma)$ and $(t,z)$ is
non-degenerate for all $\tau$ and
$\sigma$~\cite{burton_gratus_tucker:2006}.
Using (\ref{wall-of-charge_dE}) it follows that, for all $\tau>0$, $\rho$ is
well-behaved for charge distributions at rest at $\tau=0$.

The purpose of this section is to demonstrate that more general initial
conditions lead to divergences in the reduced proper charge density $\rho$ over finite
time. In~\cite{burton_gratus_tucker:2006} particular
examples of $\rho$ were generated using expressions for $V$
and $\calE$ as functions of $\tau=\hat{\tau}(t,z)$ and $\sigma=\hat{\sigma}(t,z)$.
For the present purposes it is more convenient to formulate
an ordinary differential equation for $\rho$ along $V$ and examine
properties of its solutions. The integrability condition 
\begin{equation*}
d(\rho\#\dual{V}) = 0
\end{equation*}
following from (\ref{wall-of-charge_dE}) is written
\begin{equation}
\label{equation_for_rho_along_V}
V\rho = -\rho\DivV
\end{equation}
where the scalar $\DivV\equiv\#^{-1}d\#\dual{V}$ is the divergence of
$V$. Equations (\ref{equation_for_rho_along_V}),
(\ref{equation_for_E_along_V}) and
$V\DivV=f(\rho,\DivV,\calE)$ for some $f$ forms a closed first-order ordinary
differential system for $(\rho,\DivV,\calE)$ along the integral curves
of $V$.
An explicit expression for $f$ is obtained below. 

Using the identity (see, for example, page 229 of~\cite{benn_tucker:1987})
\begin{equation}
\label{relationship_Lie_and_Del}
(\calL_V\Pg)(X,Y) = \Pg(X,\Pnabla_Y V) + \Pg(Y,\Pnabla_X V) 
\end{equation}
where $X$ and $Y$ are any vector fields on $\Pman$ and $\calL_V$ is
the Lie derivative with respect to $V$, it follows that
\begin{equation}
\label{LieV_V_as_DelV_V_working}
\begin{split}
(\calL_V\dual{V})(X) &= (\calL_V\Pg)(V,X)\\
&= \Pg(V,\Pnabla_X V) + \Pg(X,\Pnabla_V V)  
\end{split}
\end{equation}
and, using (\ref{wall-of-charge_VV}),
\begin{equation}
\label{wall-of-charge_VDelXV}
\begin{split}
\Pg(V,\Pnabla_X V) &= \frac{1}{2}\Pnabla_X\big(\Pg(V,V)\big)\\
&= 0,
\end{split}
\end{equation}
since $\Pnabla$ is metric-compatible. Therefore,
(\ref{LieV_V_as_DelV_V_working}) yields
\begin{equation}
\label{LieV_V_as_DelV_V}
\calL_V\dual{V} = \Pnabla_V\dual{V}.
\end{equation}
Using Cartan's identity~\cite{benn_tucker:1987} $\Lie_V = i_V d + d
i_V$ on forms it follows
\begin{equation}
\label{LieV_V_as_iV_dV}
\Lie_V\dual{V} = i_V d\dual{V}
\end{equation}
and so the following expressions for the relativistic acceleration
$\calA\equiv\Pnabla_V V$ of $V$ are obtained:
\begin{equation}
\label{wall-of-charge_DelVV}
\dual{\calA} = \Pnabla_V\dual{V} = i_V d\dual{V} = \calE\#\dual{V}
\end{equation}
where (\ref{wall-of-charge_lorentz}), (\ref{LieV_V_as_DelV_V}) and
(\ref{LieV_V_as_iV_dV}) have been used and
$\dual{\calA}(X)=\Pg(\calA,X)$ for any vector $X$ on $\Pman$.
Since there are no $3$-forms on $2$-dimensional manifolds
$\dual{V}\wedge d\dual{V}=0$ and
\begin{equation}
\label{dV_as_accelV}
\begin{split}
i_V(\dual{V}\wedge d\dual{V}) &= 0\\
&= (i_V\dual{V}) d\dual{V} - \dual{V}\wedge i_V d\dual{V}.
\end{split}
\end{equation}
Using (\ref{dV_as_accelV}), (\ref{wall-of-charge_DelVV}) and $i_V\dualV = \Pg(V,V) =
-1$, it follows that
\begin{equation}
\label{wall-of-charge_dV_and_hash-E}
d\dual{V} = \calE\#1.
\end{equation}
Any frame field $(X_0,X_1)$ and its dual coframe field $(e^0,e^1)$ on
$\Pman$ satisfy
\begin{align}
\notag
&e^a(X_b) = \delta^a_b,\\
\label{dual_frame_relations}
&\delta^a_b \equiv
\begin{cases}
1\quad\text{if}\,\,a=b\\
0\quad\text{if}\,\,a\neq b
\end{cases}
\end{align}
where the indices $a,b$ run over $0,1$. The intrinsic
curvature of $\Pnabla$ is zero so
\begin{equation}
\label{wall-of-charge_curvature}
\Pnabla_V\Pnabla_{X_a}V - \Pnabla_{X_a}\Pnabla_V V -
\Pnabla_{[V,X_a]}V = 0
\end{equation}
where $[V,X_a]$ is the Lie bracket of $V$ and $X_a$. The divergence
$\DivV=\#^{-1}d\#\dual{V}$ of $V$ may be written~\footnote{The Einstein
summation convention is followed i.e. repeated indices are implicitly summed.} 
\begin{equation*}
\DivV = \Pnabla\cdot V = e^a(\Pnabla_{X_a}V)
\end{equation*}
hence
\begin{equation}
\label{equation_for_DivV_derivation_1}
\begin{split}
V\DivV &= \Pnabla_V(\Pnabla\cdot V)\\
&= (\Pnabla_V e^a)(\Pnabla_{X_a}V) + e^a(\Pnabla_V\Pnabla_{X_a}V) \\
&= (\Pnabla_V e^a)(\Pnabla_{X_a}V) +
\Pnabla\cdot\calA + e^a(\Pnabla_{[V,X_a]}V) 
\end{split}
\end{equation}
where (\ref{wall-of-charge_curvature}) and
(\ref{wall-of-charge_DelVV}) have been used.
The torsion of $\Pnabla$ vanishes
\begin{equation*}
\Pnabla_V X_a - \Pnabla_{X_a} V - [V,X_a] = 0
\end{equation*}
and so, using (\ref{dual_frame_relations})
\begin{equation}
\begin{split}
\Pnabla_V e^a &= -\big(e^a(\Pnabla_V X_b)\big)\,e^b\\
&= -\big(e^a(\Pnabla_{X_b}V) + e^a([V,X_b])\big)\,e^b.
\end{split}
\label{DelV_e}
\end{equation}
Using (\ref{DelV_e}) to eliminate $\Pnabla_V e^a$ in (\ref{equation_for_DivV_derivation_1}) yields
\begin{equation}
\label{equation_for_DivV_derivation_2}
V\DivV = \Pnabla\cdot\calA - \text{tr}(\Pnabla V \Pnabla V) 
\end{equation}
where $\Pnabla V = e^b(\Pnabla_{X_a}V)\, e^a\Tensor X_b$
and $\text{tr}(\Pnabla V \Pnabla V) =
e^b(\Pnabla_{X_a}V)e^a(\Pnabla_{X_b}V)$. The scalar $\text{tr}(\Pnabla
V \Pnabla V)$ is obtained using the following $\Pg$-orthonormal frame field
$\{X_0,X_1\}$ adapted to $V$ and its dual frame field $\{e^0,e^1\}$:
\begin{align}
\notag
&X_0 = V,\quad X_1 = \dual{\#\dual{V}},\\
\label{wall-of-charge_frame_field}
&e^0 = -\dual{V},\quad e^1 = \#\dual{V}
\end{align}
where the vector field $X_1=\dual{\#\dual{V}}$ is defined by
$\Pg(X,X_1) = (\#\dual{V})(X)$ on any vector $X$ on $\Pman$.
Using (\ref{wall-of-charge_VDelXV}), (\ref{relationship_Lie_and_Del})
and (\ref{wall-of-charge_frame_field}) it follows that
\begin{equation}
\label{trDelVDelV_as_LieV-Pg}
\begin{split}
\text{tr}(\Pnabla V \Pnabla V) &= \big(\Pg(X_1,\Pnabla_{X_1}V)\big)^2\\
&= \frac{1}{4}\big((\Lie_V\Pg)(X_1,X_1)\big)^2
\end{split}
\end{equation}
and expressing $\Pg$ as
\begin{equation*}
\begin{split}
\Pg &= -e^0\Tensor e^0 + e^1\Tensor e^1\\
&= -\dual{V}\Tensor\dual{V}+\#\dual{V}\Tensor\#\dual{V}
\end{split}
\end{equation*}
it follows 
\begin{equation*}
\begin{split}
\Lie_V\Pg = &-\Lie_V\dual{V}\Tensor\dual{V} +
\dual{V}\Tensor\Lie_V\dual{V}\\
&+ \Lie_V\#\dual{V}\Tensor
\#\dual{V}+\#\dual{V}\Tensor \Lie_V\#\dual{V}
\end{split}
\end{equation*}
and
\begin{equation}
\label{LieV-Pg_as_DivV}
\begin{split}
(\Lie_V\Pg)(X_1,X_1) &= 2(\Lie_V\#\dual{V})(X_1)\\
&= 2\DivV
\end{split} 
\end{equation}
where (\ref{dual_frame_relations}),
(\ref{wall-of-charge_frame_field}), $\DivV=\#^{-1}d\#\dual{V}$ and Cartan's identity $\Lie_V
= i_V d + d i_V$ on forms have been used. Equations (\ref{equation_for_DivV_derivation_2}),
(\ref{LieV-Pg_as_DivV}) and (\ref{trDelVDelV_as_LieV-Pg}) give
\begin{equation}
\label{equation_for_DivV_derivation_3}
V\DivV = \Pnabla\cdot\calA - \DivV^2
\end{equation}
and writing $\Pnabla\cdot\calA$ as a differential form yields
\begin{equation}
\label{DivA_opened}
\begin{split}
\Pnabla\cdot\calA &= \#^{-1}d\#\dual{\calA}\\
&= \#^{-1}d\# i_V d\dual{V}\\
&= -\#^{-1}d(\dual{V}\wedge\# d\dual{V})\\
&= -\#^{-1}(d\dual{V}\wedge\# d\dual{V}) + \#^{-1}(\dual{V}\wedge d\#
d\dual{V}).
\end{split}
\end{equation}
Thus, using (\ref{wall-of-charge_dE}),
(\ref{wall-of-charge_dV_and_hash-E}),
and (\ref{DivA_opened}) it
follows that (\ref{equation_for_DivV_derivation_3}) is
\begin{equation}
\label{equation_for_DivV_along_V}
V\DivV = \calE^2 + \rho - \DivV^2.
\end{equation}
Equations (\ref{equation_for_DivV_along_V}),
(\ref{equation_for_rho_along_V}) and (\ref{equation_for_E_along_V})
are a closed system of differential equations for $(\rho,\DivV,\calE)$
along $V$.

Let $\myC$ be any proper-time parametrised integral curve of $V$:
\begin{align*}
\begin{split}
\myC : I &\rightarrow \Pman,\\
\lambda &\rightarrow \big( t=T(\lambda), z=Z(\lambda)\big)
\end{split}
\end{align*}
where $I$ is a subset of the real line $\mathbb{R}$ and
\begin{align*}
\Gamma_*\PD_\lambda = V.
\end{align*} 
Hence, equations (\ref{equation_for_E_along_V}),
(\ref{equation_for_rho_along_V}) and (\ref{equation_for_DivV_along_V})
pulled back to $\mathbb{R}$ using $\myC^*$ are 
\begin{align}
\label{ODE_for_E_C}
&\frac{d\calE_\myC}{d\lambda} = 0,\\
\label{ODE_for_rho_C}
&\frac{d\rho_\myC}{d\lambda} = -\rho_\myC\DivV_\myC,\\
\label{ODE_for_DivV_C}
&\frac{d\DivV_\myC}{d\lambda} = \calE_\myC^2 + \rho_\myC - \DivV_\myC^2
\end{align}
where the subscript $\myC$ indicates pull-back using $\myC^*$
e.g. $\calE_\myC(\lambda) = (\myC^*\calE)(\lambda) =
\calE\big(T(\lambda),Z(\lambda)\big)$.
The general solution to (\ref{ODE_for_rho_C}) is
\begin{equation}
\label{rho_C_is_positive}
\rho_\myC(\lambda) =
\rho_0\exp\bigg(-\int^\lambda_0\DivV_\myC(\lambda^\prime)
d\lambda^\prime\bigg)
\end{equation}
where $\rho_0=\rho_\myC(0)$ is a value of $\rho$ on an initial
hypersurface. Since $\rho=\frac{q_0^2}{\epsilon_0 m_0 c^2}\calN$ and
the proper number density $\calN \ge 0$ it follows $\rho_\myC\ge 0$.
Over intervals of $\lambda$ on
which $d\rho_\myC/d\lambda \neq 0$
(\ref{ODE_for_E_C}-\ref{ODE_for_DivV_C}) yields
\begin{align*}
&\frac{d\calE_\myC}{d\rho_\myC} = 0,\\
&\frac{d\DivV_\myC}{d\rho_\myC} = \frac{1}{-\rho_\myC\DivV_\myC}(\calE_\myC^2+\rho_\myC-\DivV_\myC^2)
\end{align*}
and so 
\begin{equation*}
\rho^2_\myC\frac{d}{d\rho_\myC}(\rho^{-2}_\myC\DivV^2_\myC) =
-\frac{2\calE^2_\myC}{\rho_\myC}-2 
\end{equation*}
leading to the first integral
\begin{equation}
\label{ODEs_first_integral}
\DivV_\myC^2 = \calE_0^2 + 2\rho_\myC + \kappa_0\rho_\myC^2 
\end{equation}
of (\ref{ODE_for_E_C}-\ref{ODE_for_DivV_C}) where $\kappa_0$ is a constant of
integration determined by the initial values $\DivV_\myC(0)=\DivV_0$,
$\rho_\myC(0)=\rho_0$ and $\calE_\myC(0)=\calE_0$.

According to (\ref{ODEs_first_integral}) $\DivV_\myC^2$ is a quadratic function
in $\rho_\myC$ with $\calE_0$ and $\kappa_0$ held fixed and
the large $\lambda$ behaviour of $\rho_\myC$ and
$\DivV_\myC$ crucially depend on the sign of $\kappa_0$. Since
$\calE_0^2$ is a positive constant and $\DivV_\myC^2$ is positive
($\DivV_\myC$ is real), it follows that if $\kappa_0<0$ then $\rho_\myC$ in
(\ref{ODEs_first_integral}) cannot be arbitrarily large. 
\begin{equation}
\label{negative_kappa_rho_bounded}
\text{\emph{If $\kappa_0<0$ then $\rho_\myC$ is bounded from above.}}
\end{equation}
However, if $\kappa_0\ge 0$ then no such
bound on $\rho_\myC$ exists and, in principle, $\rho_\myC$ and
$\DivV_\myC$ can attain arbitrarily large values.
In fact, as will now be shown, if
$\kappa_0 > 0$ then $\rho_\myC$ may diverge in \emph{finite}
proper time.

Assume that the initial data satisfies $\kappa_0>0$ and $\DivV_0<0$.
Using (\ref{ODEs_first_integral}) to eliminate $\DivV_\myC^2$ from the
right-hand side of (\ref{ODE_for_DivV_C}) leads to
\begin{equation}
\label{DivVprime_rho}
\frac{d\DivV_\myC}{d\lambda} = - \rho_\myC - \kappa_0\rho_\myC^2
\end{equation}
and since $\rho_\myC\ge 0$, $\kappa_0>0$ and $\DivV_0<0$ it follows
from (\ref{DivVprime_rho}) that
$\DivV_\myC<0$. Therefore, using the \emph{negative} root of (\ref{ODEs_first_integral}),
\begin{equation*}
\DivV_\myC = -\sqrt{\calE_0^2 + 2\rho_\myC + \kappa_0\rho_\myC^2},
\end{equation*}
to eliminate $\DivV_\myC$ from the right-hand side of
(\ref{ODE_for_rho_C})
\begin{equation*}
\frac{d\rho_\myC}{d\lambda} = \rho_\myC\sqrt{\calE_0^2 + 2\rho_\myC + \kappa_0\rho_\myC^2}
\end{equation*}
is obtained and $\rho_\myC$ asymptotes at proper time $\lambda_\infty$ where
\begin{equation}
\label{proper_time_interval}
\lambda_\infty = \int\limits^\infty_{\rho_0}\frac{1}{\rho_\myC\sqrt{\calE_0^2 + 2\rho_\myC + \kappa_0\rho_\myC^2}}\,d\rho_\myC.
\end{equation}
The integrand is positive and $(\calE_0^2 + 2\rho_\myC)>0$ so
(\ref{proper_time_interval}) implies
\begin{equation*}
\begin{split}
\lambda_\infty &<
\int\limits^\infty_{\rho_0}\frac{1}{\rho_\myC\sqrt{\kappa_0\rho_\myC^2}}\,d\rho_\myC\\
&= \frac{1}{\rho_0\sqrt{\kappa_0}}
\end{split}
\end{equation*}
i.e. $\lambda_\infty$ is bounded from above.
\begin{equation}
\label{rho_diverges_over_finite_deltalambda}
\begin{split}
&\text{\emph{If $\kappa_0>0$ and $\DivV_0<0$ at $\lambda=0$ then
$\rho_\myC$ diverges}}\\
&\text{\emph{at proper time $\lambda=\lambda_\infty$ less than
$1/(\rho_0\sqrt{\kappa_0})$.}} 
\end{split}
\end{equation}
The constant $\kappa_0$ may be obtained in terms of data on the
spacelike hypersurface $t=0=T(0)$ where $(t,z)$ is the laboratory
coordinate system.
Let
$\mu(0,z)$ be the initial Newtonian velocity measured in the
laboratory frame (see (\ref{wall-of-charge_ansatz}) for the definition
of $\mu$). Using
(\ref{wall-of-charge_ansatz}), $\DivV=\Pnabla\cdot V$ and (\ref{wall-of-charge_lorentz}) it follows that
\begin{align}
\label{coordinate_expression_DivV}
&\DivV = \partial_t\gamma + \partial_z(\gamma\mu),\\
\label{coordinate_expression_Lorentz}
&\partial_t\gamma + \mu\partial_z\gamma = \calE\mu
\end{align}
where $\gamma=1/\sqrt{1-\mu^2}$. Using
(\ref{coordinate_expression_Lorentz}) to eliminate $\partial_t\gamma$
from (\ref{coordinate_expression_DivV}) yields
\begin{equation*}
\DivV = \calE\mu + \gamma\partial_z\mu
\end{equation*}
and so on the initial spacelike hypersurface $t=0$
\begin{equation}
\label{initial_DivV}
\DivV_0 = \calE_0\mu_0 + \gamma_0(\partial_z\mu)_0
\end{equation}
where $\mu_0=(\myC^*\mu)(0)$,
$(\partial_z\mu)_0=\big(\myC^*(\partial_z\mu)\big)(0)$. Using
(\ref{initial_DivV}) to eliminate $\DivV_0=\DivV_\myC(0)$ in
(\ref{ODEs_first_integral}) evaluated at $\lambda=0$ gives 
\begin{equation}
\label{value_of_kappa0}
\kappa_0 = \frac{1}{\rho_0^2}\bigg(\gamma_0^2(\partial_z\mu)_0^2 +
2\calE_0\mu_0\gamma_0(\partial_z\mu)_0-\frac{\calE_0^2}{\gamma_0^2}-2\rho_0\bigg).
\end{equation}
Equation (\ref{value_of_kappa0}) indicates that if
$(\partial_z\mu)_0=0$ then $\kappa_0<0$ and according to
(\ref{negative_kappa_rho_bounded}) $\rho_\myC$ does not diverge.
This result agrees with the non-singular behaviour
of the exact solutions to
(\ref{wall-of-charge_dE}-\ref{wall-of-charge_VV}) satisfying
$\mu(0,z)=0$  presented in~\cite{burton_gratus_tucker:2006}; 
for $\kappa_0$ to be positive $(\partial_z\mu)_0$ must be non-zero.

\begin{figure}[h]
\begin{center}
\scalebox{0.5}{\includegraphics{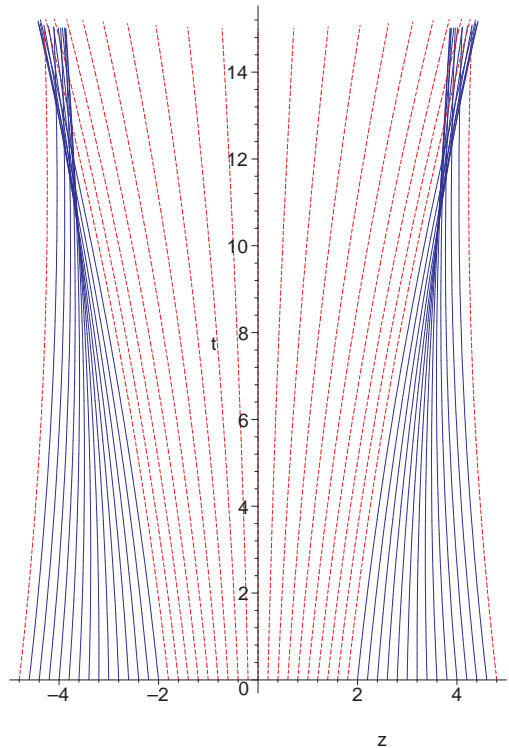}}
\caption{\label{figure:tz_crossings}The integral curves of $V$ for a
particular solution to equations
(\ref{wall-of-charge_dE}-\ref{wall-of-charge_VV}).}
\end{center}
\end{figure}

\begin{figure}[h]
\begin{center}
\scalebox{1.0}{\includegraphics{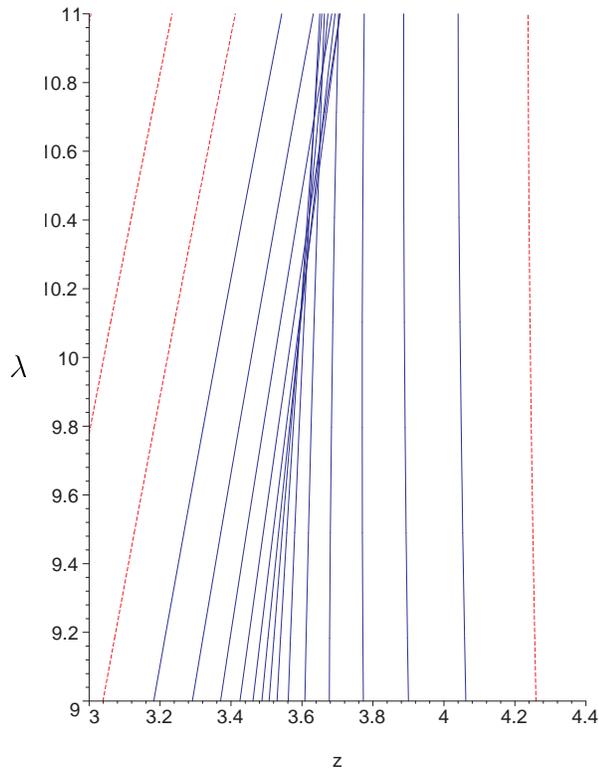}}
\caption{\label{figure:propertz_crossings}The integral curves of $V$ for a
particular solution to equations
(\ref{wall-of-charge_dE}-\ref{wall-of-charge_VV}).}
\end{center}
\end{figure}

The integral curves of $V$ for the particular solution to equations
(\ref{wall-of-charge_dE}-\ref{wall-of-charge_VV}) with the initial conditions
\begin{align*}
&\calE(0,z) = \frac{1}{2}\bigg(\int\limits_{-\infty}^z
\frac{\rho(0,s)}{\sqrt{1-\mu(0,s)^2}}\,ds -
\int\limits_z^\infty\frac{\rho(0,s)}{\sqrt{1-\mu(0,s)^2}}\,ds
\bigg),\\
&\rho(0,z) = 0.01\exp(-z^2),\\
&\mu(0,z) = 0.1\sin(z)
\end{align*}
are shown in figure~\ref{figure:tz_crossings}. Trajectories on which
$\kappa_0>0$ and $\DivV_0<0$ are solid and trajectories on
which $\kappa_0<0$ are dashed. Thus,
according to (\ref{rho_diverges_over_finite_deltalambda}), $\rho$
diverges at points along the solid trajectories, which meet
at about $t=11$. The details of the onset of the crossings are shown in
figure~\ref{figure:propertz_crossings} with proper time $\lambda$ and
$z$ as axes. Evaluating (\ref{proper_time_interval}) gives
$\lambda_\infty=10.14$ for the trajectory starting at
$(t,z)=(0,3)$. 
Any comoving coordinate system $(\tau,\sigma)$ adapted to the
trajectories, where $\sigma$ is constant on each trajectory,
degenerates where the solid trajectories cross. The
determinant of the Jacobian of the $(\tau,\sigma)\rightarrow (t,z)$
transformation vanishes at such points and $\rho$ diverges.
\section{Discussion of crossing trajectories in charged particle beams}
\label{section:discussion_of_crossing_trajectories}
In fluid and gas dynamics, crossing trajectories are considered to be a symptom of
incomplete physics. All real fluids and gases are viscous to some extent
and trajectories may cross if their viscosity is neglected. For
example, a compression wave in a hypothetical inviscid fluid may lead
to a velocity field with crossing characteristics, i.e. the velocity becomes
multi-valued.
In reality this is not what happens; the velocity remains
single-valued and stabilises to form a propagating shock. The fluid on
either side of the shock is essentially inviscid but close to the
shock the second-order spatial derivatives of the velocity are so
large that dissipation can no longer be neglected.

However, it is far from clear that such arguments are relevant for a charged
particle beam, which is physically very different from a normal
fluid. Although \Fieldsysrefs are often said to describe a ``cold
charged fluid''~\cite{davidson:2001}, this terminology is misleading.
Microscopically, a normal fluid is a complicated system
of neutral particles whose interactions are dominated by molecular
collisions and possibly gravity but the dominant inter-particle forces in a
beam of electrons are entirely electromagnetic in origin.
At first sight, it seems that all of the necessary physics is
contained in \Fieldsysrefsnotilde.

The velocity field $V$ is a smoothed out representation of the particle
motion. The fact that the trajectories cross does not mean that the
particles are colliding; it merely indicates that a smooth field representation
of the particles has degenerated. Similarly, the $3$-volume number
density of a collection of particles may attain arbitrarily large
values if the particles dynamically arrange themselves into planar or
linear configurations.

Nevertheless, the electric field induced by $V$ in
\Fieldsysrefs is inconsistent where trajectories of $V$ cross. Equation
(\ref{equation_for_E_along_V}) indicates that the electric field in any wall-of-charge solution is constant along
$V$. However, during the crossing the charge distribution ``passes
through itself'' and the electric field has to change. The reason for
this is easiest to appreciate by the following simple analogy. Consider the dynamics of a pair of positive \emph{sheet} charges that are permitted to pass
through each other. The sheets are arranged so that their normals and
their electric fields lie along the $z$-axis. Each sheet has the same
properties; its self-induced electric field is $E$ at all points to its right and
$-E$ at all points its left, where $E>0$ is constant, and its charge per unit
area is $Q=2\epsilon_0 E$.
The sheets are labelled $1$ and $2$ and
sheet $1$ lies to the left of sheet $2$ initially. The electrostatic
force acting on each sheet is constant and is due to the electric
field of the other sheet. The force per unit area acting on sheet $1$
is $-QE$ and on sheet $2$ is $QE$. Thus, the sheets will repel each other
but their electric forces will not be enough to stop them meeting for
sufficiently large opposing initial velocities. Let the initial
velocities be so large that the sheets pass through each
other. At all times before they meet, the force on sheet $1$ is $-QE$
and the force on sheet $2$ is $QE$. After they meet their
r\^oles have been exchanged; sheet $2$ is acted on by $-QE$
while sheet $1$ is acted on by $QE$. Now consider a large number of
sheets undergoing collective motions in which only \emph{some} of the
sheets pass through each other. A continuum realisation of this model
is a dynamical set of component continua each with its own
velocity field. The number of components evolves in time and depends
on the history of the total continuum.    

The spacetime fields $(V,\rho)$ satisfying \Fieldsysrefs offer an Eulerian
description of a single component continuum. Although a dynamical
number of components can be simulated using more complicated Eulerian field
theories~\cite{li_wohlbier_etal:2004}, such approaches are restrictive
because an upper bound must be placed on the number of anticipated
components. In the following section
a new Lagrangian model of a multi-component charged continuum is
proposed in which the number of components is dynamical and free to attain
\emph{any} value. The essential idea is to describe the bulk particle motion
using a flow map $C$ from a body-time manifold into spacetime rather
than inducing the motion from a velocity field $V$ on spacetime. $C$ may be
described as ``folding'' a single electric current on the body-time
manifold to give a multi-component current on spacetime.   
\section{Lagrangian description of multi-component charged continua}
\label{section:lagrangian_cold_charge}
Ingredients in the following Lagrangian description are an auxiliary
$4$-dimensional manifold $\Bman$ called the \emph{body-time} manifold and a
map $C$ from $\Bman$ into Minkowski spacetime $\man$.
The body-time manifold
$\Bman=\Real\times\UBman$ where $\UBman$ is a $3$-dimensional
\emph{body} manifold so each point $P\in\Bman$ is also written
$P=(\lambda,\UP)$ for $\lambda\in\Real$ and $\UP\in\UBman$. Each point
$\UP\in\UBman$ generates a curve $C_\UP$ in spacetime $\man$ where
\begin{equation*}
C_\UP(\lambda) = C(\lambda,\UP).
\end{equation*}
The map $C$ is normalised so that $\lambda$ is the proper-time parameter
of $C_\UP$ for all $\UP\in\UBman$: 
\begin{equation}
\label{multi_C_normalisation}
g(\dot{C},\dot{C})=-1
\end{equation}
where $\dot{C}(P)=(C_*\partial_\lambda)(P)$ is a vector at
$p=C(P)$ in $\man$. 

In general, $C$ is a many-to-one
map, i.e. there exists $P_1$ and $P_2$ in $\Bman$ such that
$C(P_1)=C(P_2)$, and $C$ is not required to be surjective.
For any point $p\in\man$ there may exist any number $N(p)$
of real roots of the equation $p=C(P)$. The map $C$ describes
``multi-valued velocities'' because although
$C(P_1)=C(P_2)$, there is no reason why $\dot{C}(P_1)$
should equal $\dot{C}(P_2)$. Thus, in general $\dot{C}$ cannot be
identified with a vector field on $\man$. The domain of $\dot{C}$ is
$\Bman$ and $\dot{C}(P)$ is a vector at the point $C(P)$;
the map $\dot{C}$ is referred to as a vector field over $C$.
 
The map $C$ is defined to satisfy the Lorentz force equation
\begin{equation}
\label{multi_lorentz}
\nabla_{\dot{C}}\widetilde{\dot{C}} = i_{\dot{C}}F
\end{equation}
where $F$ is an electromagnetic field $2$-form on
$\man$ and $(\nabla_{\dot{C}}\widetilde{\dot{C}})(P)$ and
$i_{\dot{C}(P)}F(p)$
are covectors at $p=C(P)$. The maps
$\nabla_{\dot{C}}\widetilde{\dot{C}}$ and $i_{\dot{C}}F$ are
covector fields over $C$ (i.e. $1$-forms over $C$).

The set inverse $C^{-1}$ of $C$ at $p$ includes the set of points in
$\Bman$ for which $p=C(P)$ and is written 
\begin{equation*}
C^{-1}(\{p\}) =
\begin{cases}
\strut\{P_{[1]},P_{[2]},\dots,P_{[N(p)]}\strut\}\quad\text{if $N(p)\ge
1$}\\
\emptyset\quad\text{if $N(p)=0$}
\end{cases}
\end{equation*}
where
the square brackets distinguish root labels from coordinate and frame
labels.     

Each element of $C^{-1}(\{p\})$ gives rise to a \emph{partial}
electric $4$-current $J_{[i]}(p)$, where $i=1,2,...,N(p)$. The sum of
partial currents is the total electric $4$-current driving $F$ in the
Maxwell equations   
\begin{align}
\label{multi_faraday_no-magnetic-monopoles}
&dF = 0,\\
\label{multi_gauss_ampere-maxwell}
&d\star F = -\sum\limits^{N(p)}_{i=0}\star\dual{J_{[i]}}.
\end{align}

Regions of spacetime with different numbers of partial currents are
distinguished by examining the pull-back $C^*(\star 1)$ of the spacetime volume
$4$-form $\star 1$ by $C$. \emph{Critical} points in $\Bman$ are
defined by the vanishing of the $4$-form $C^*(\star 1)$. Their
images under $C$ are also said to be critical and lie in the
interfaces between spacetime regions with different $N(p)$. Specifying
a non-vanishing closed $3$-form $\calJ$ on $\Bman$ satisfying 
\begin{equation}
\label{multi_basic_form}
i_{\PD_\lambda}\calJ = 0,\quad\Lie_{\PD_\lambda}\calJ = 0
\end{equation}
leads to a scalar field $\Delta$ on $\Bman$ where
\begin{equation}
\label{multi_Delta}
\Delta d\lambda\wedge\calJ = C^*(\star 1)
\end{equation}
which vanishes at critical points. At each point $p$ in $\man$ the partial
$4$-current $J_{[i]}(p)$ has the form
\begin{equation}
\label{multi_partial_currents}
J_{[i]}(p) = \Brho(P_{[i]})\dot{C}(P_{[i]})\quad\text{where
$P_{[i]}=C^{-1}_{[i]}(p)$}
\end{equation}
and $\Brho$ is the scalar field
\begin{equation}
\label{multi_proper_charge_density}
\Brho=\frac{1}{|\Delta|}
\end{equation}
on $\calB$.
The system \Multisysrefs differs significantly from \Fieldsysrefs
because the number $N(p)$ of elements of $C^{-1}(\{p\})$, and
therefore the number of partial $4$-currents in
(\ref{multi_gauss_ampere-maxwell}), depends on the spacetime point
$p$. Numerically integrating \Multisysrefs involves computing $N(p)$
at each time step.
\section{Multi-component charge configurations}
\label{section:multi_component_charge}
\begin{figure}[h]
\begin{center}
\scalebox{0.5}{\includegraphics{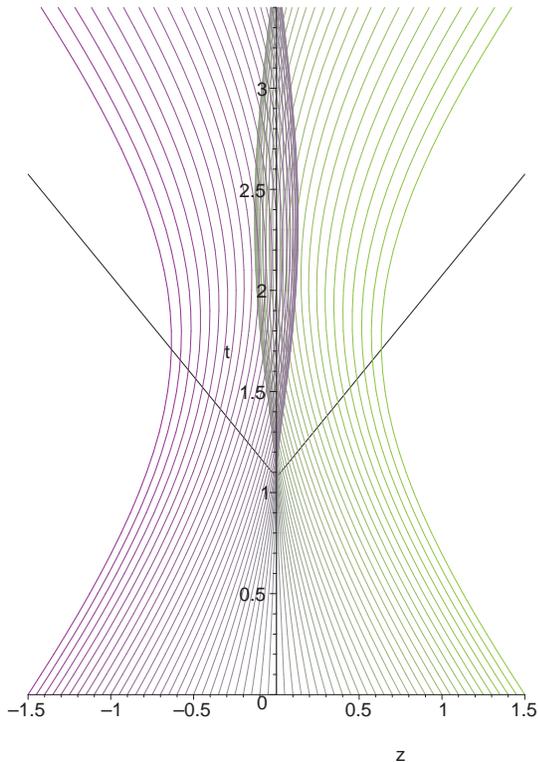}}
\caption{\label{figure:single_component_trajectories}The trajectories
of the particular solution to \Fieldsysrefs with the initial
conditions
(\ref{multi_current_example_initial_conditions}). Trajectories cross
in a narrow region inside the forward light-cone of the critical point
$p_\text{crit}=(0,1.075)$.} 
\end{center}
\end{figure}

\begin{figure}[h]
\begin{center}
\scalebox{0.5}{\includegraphics{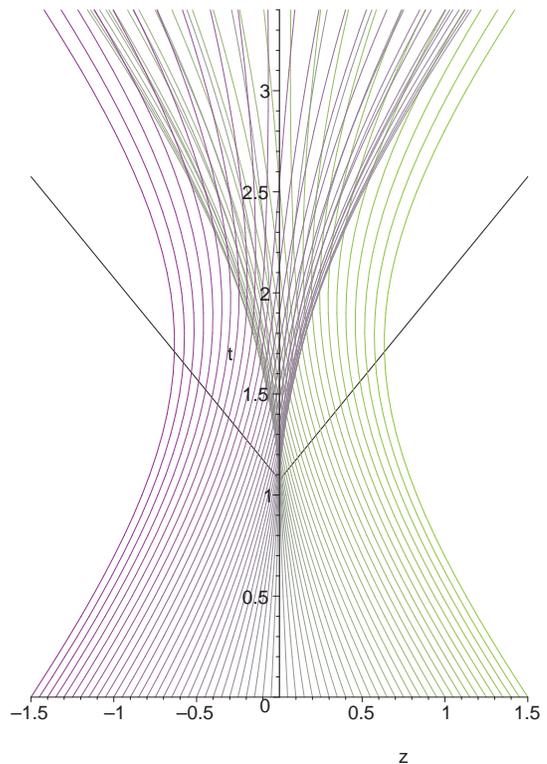}}
\caption{\label{figure:multi_component_trajectories}The trajectories
of the particular solution to \Multisysrefs with the initial
conditions (\ref{multi_current_example_initial_conditions}).
The forward light-cone of the critical point $p_\text{crit}=(0,1.075)$
is also shown. The region outside the ``fan'' emanating from
$p_\text{crit}$ has only one partial current whereas the region inside the
``fan'' contains three partial currents.
}
\end{center}
\end{figure}

Since \Fieldsysrefs are equivalent to \Multisysrefs when applied to a spacetime
region with a single partial current, it follows that solutions to
\Fieldsysrefs and \Multisysrefs agree in such a region. This is illustrated in
figures~\ref{figure:single_component_trajectories}
and~\ref{figure:multi_component_trajectories} by collapsing ``wall of
charge'' solutions to \Fieldsysrefs and \Multisysrefsnotilde. The
ans\"atze for $F$ and $V$ are
\begin{align*}
&F=\calE(t,z) dt\wedge dz,\\
&V=\cosh\chi(t,z) \PD_t + \sinh\chi(t,z) \PD_z
\end{align*}
and $\rho$ depends only on $(t,z)$ where $(t,x,y,z)$ is an inertial
Cartesian coordinate system with 
\begin{equation*}
g = -dt\Tensor dt + dx\Tensor dx + dy\Tensor dy + dz\Tensor dz.
\end{equation*}
Similarly, $C$ and $\Brho$ only depend on $(\lambda,\sigma^1)$ where
$(\sigma^1,\sigma^2,\sigma^3)$ is a coordinate system on $\UBman$.
The initial conditions on $\calE$, $\chi$ and $\rho$ are
\begin{align}
\notag
&\calE(0,z) = \frac{1}{2}\bigg(\int\limits_{-\infty}^z
(\rho\cosh\chi)(0,s)\,ds -
\int\limits_z^\infty(\rho\cosh\chi)(0,s)\,ds
\bigg),\\
\notag
&\rho(0,z) =
\begin{cases}
 0.025\qquad&\text{if $-1.5\le z\le 1.5$}\\
0\qquad&\text{otherwise},
\end{cases}\\
\label{multi_current_example_initial_conditions}
&\chi(0,z) = -1.2\tanh(z)
\end{align}
with analogous initial conditions on $C$ and $\Brho$.

In both cases
a critical point develops at $p_\text{crit}=(0,1.075)$.
The integral curves of $V$ in the single-component Eulerian
model \Fieldsysrefs exhibit crossings along a narrow corridor inside
the forward light-cone of $p_\text{crit}$ (see
figure~\ref{figure:single_component_trajectories}). However, the
solution to \Multisysrefs shown in
figure~\ref{figure:multi_component_trajectories} contains a ``fan'' of
three partial currents inside $p_\text{crit}\text{'s}$ forward light-cone.
The two models clearly yield dramatically different results.

On the other hand, the two models agree outside the forward light-cone at
$p_\text{crit}$. An argument for this is as follows: 
Integrate \Fieldsysrefs and \Multisysrefs using a
time slicing adapted to a field of synchronised inertial observers
moving along the $z$-axis in the laboratory frame with positive
constant velocity arbitrarily close to the speed of light. As before, the solutions
agree up to a constant proper time surface containing the point $p_\text{crit}$.
Part of this proper time hypersurface almost coincides with the $z>0$ subset
of $p_\text{crit}\text{'s}$ forward light cone. Now integrate
the equations using a time-slicing adapted to a field of synchronised
inertial observers moving along the $z$-axis in the negative direction
at almost the speed of light. The result agrees almost up to the $z<0$
subset of $p_\text{crit}\text{'s}$ forward light-cone. It follows that
the solutions to
\Fieldsysrefs and \Multisysrefs agree at points outside the
forward light-cone at $p_\text{crit}$. 

\begin{figure}[h]
\begin{center}
\scalebox{1.0}{\includegraphics{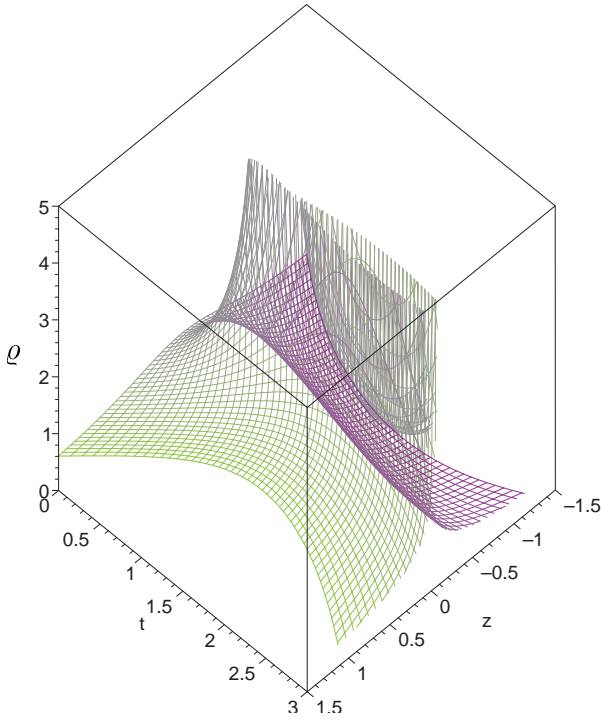}}
\caption{\label{figure:single_component_charge}The reduced proper
charge density of the particular solution to \Fieldsysrefs with the initial
conditions (\ref{multi_current_example_initial_conditions}).}
\end{center}
\end{figure}

\begin{figure}[h]
\begin{center}
\scalebox{1.0}{\includegraphics{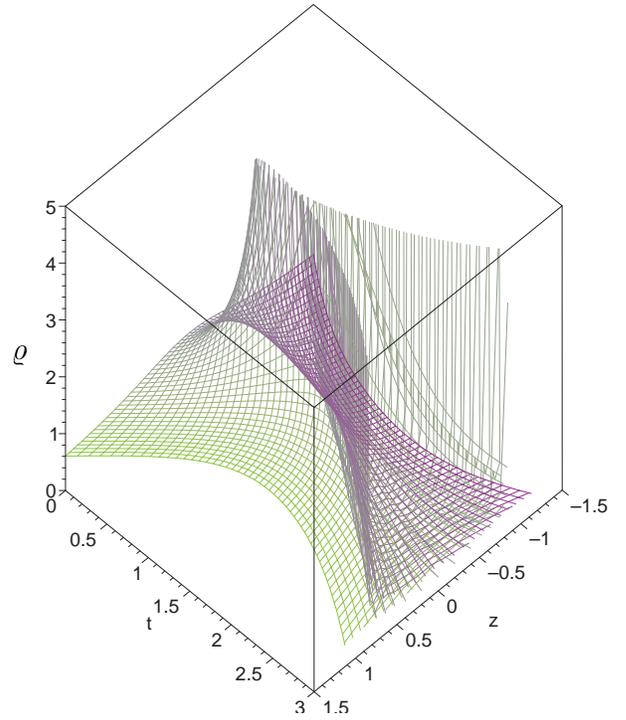}}
\caption{\label{figure:multi_component_charge}The partial reduced proper
charge densities of the particular solution to \Multisysrefs with the initial
conditions (\ref{multi_current_example_initial_conditions}). The
region outside the ``fan'' has only one partial current
whereas the region inside the ``fan'' contains three partial
currents.
}
\end{center}
\end{figure}

For further comparison the
reduced proper charge densities for \Fieldsysrefs and \Multisysrefs are shown in
figures~\ref{figure:single_component_charge}
and~\ref{figure:multi_component_charge}.  
\section{Ultra-relativistic approximation scheme}
\label{section:ultra-relativistic_approximation}
Ultra-relativistic solutions to \Fieldsysrefs may be
obtained by promoting $(V,\rho,F)$ to a $1$-parameter family
$(V^\Vep,\rho^\Vep,F^\Vep)$ in $\Vep$ where the $\Vep$ dependences are
motivated by exact ``wall of charge''
solutions~\cite{burton_gratus_tucker:2006}. The equations
obtained by equating orders in $\Vep$ lead
to a self-consistent hierarchical method for approximating solutions to
\Fieldsysrefsnotilde. The virtue of the scheme is
that the $\Vep$ dependences conspire to produce an infinite tower of
equations that are partially coupled and
are generally easier to solve than the fully coupled system
\Fieldsysrefsnotilde.

A similar approach based on \Multisysrefs will now be outlined.
Let $C^\Vep$ be a $1$-parameter family of maps from $\Bman$ into
$\man$ such that
\begin{equation*}
g(\dot{C}^\Vep,\dot{C}^\Vep) = -1
\end{equation*}
where $\dot{C}^\Vep = C^\Vep_*\PD_\lambda$. Let $\lambda=\varepsilon
s$ and introduce the map $\VarC^\Vep$ where
\begin{equation}
\label{multi_VarC_and_C}
\VarC^\Vep(s,\UP) = C^\Vep(\varepsilon s,\UP)
\end{equation}
at all non-critical points in the domain of $C^\Vep$. Thus
\begin{equation}
\label{multi_change_of_variable}
\dot{C}^\Vep = \frac{1}{\Vep}\VarC^\Vep_*\PD_s =
\frac{1}{\Vep}\VarCdot{\Vep} 
\end{equation}
where $\VarCdot{\Vep} = \VarC^\Vep_*\PD_s$ and so
\begin{equation}
\label{multi_expansion_C_normalisation}
g(\VarCdot{\Vep},\VarCdot{\Vep}) = -\Vep^2.
\end{equation}
and the Lorentz force equation for $\VarC^\Vep$ becomes
\begin{equation}
\label{multi_lorentz_force_epsilon}
\nabla_{\VarC^{\Vep\prime}}\dual{\VarC^{\Vep\prime}}
=\Vep i_{\VarC^{\Vep\prime}}F^\Vep.
\end{equation}

In~\cite{burton_gratus_tucker:2006} a dependence for $V^\Vep$ on
$\Vep$ of the form 
\begin{equation}
\label{ansatz_for_V_Vep}
V^\Vep = \sum^\infty_{n=-1} \Vep^n V_n
\end{equation}
was motivated by ``wall of charge'' solutions to
\Fieldsysrefsnotilde. For multi-current configurations on Minkowski
spacetime $\man$, the flow map $\VarC$ is the dependent variable and
it is natural to exploit the affine structure of
$\man$ and postulate an analogous series for $\VarC^\Vep$.
Let $(x^a)$ be an inertial Cartesian
coordinate system adapted to the laboratory frame on $\man$ where
$a,b=0,1,2,3$ and 
\begin{equation*}
g = \eta_{ab}dx^a\Tensor dx^b
\end{equation*}
where
\begin{equation*}
\eta_{ab} =
\begin{cases}
-1\qquad&\text{if $a=b=0$}\\
1\qquad&\text{if $a=b\neq 0$}\\
0\qquad&\text{if $a\neq b$.}
\end{cases}
\end{equation*}
In the rest of this article the map $\VarC^\Vep$ is also regarded as the
$4$-component column vector
\begin{equation*}
\left(
\begin{array}{c}
\VarC^{\Vep 0}\\
\VarC^{\Vep 1}\\
\VarC^{\Vep 2}\\
\VarC^{\Vep 3}
\end{array}
\right)
\end{equation*}
where $(\VarC^{\Vep a})$ are the $(x^a)$ components of $\VarC^\Vep$. 

The $\Vep$ expansions of $\VarC^{\Vep a}(Q)$, where $Q=(s,\UP)$, are chosen as
\begin{equation}
\label{C_epsilon_expansion}
\VarC^{\Vep a}(Q) = \sum_{n=0}^\infty \varepsilon^n \VarC^a_n(Q).
\end{equation}
Motivated by the corresponding expression in the Eulerian
single-current formulation~\cite{burton_gratus_tucker:2006}, the
$1$-parameter family $F^\Vep$ of electromagnetic $2$-forms is
chosen as
\begin{equation}
\label{F_epsilon_expansion}
F^\Vep(p) = \sum_{n=-1}^\infty \varepsilon^n F_n(p)
\end{equation}
at any point $p$ in $\man$. Note that (\ref{F_epsilon_expansion}) is
independent of the $\Vep$ expansion of $\VarC^\Vep$ in
(\ref{C_epsilon_expansion}); $F^\Vep$ is a $1$-parameter family of
$2$-forms on Minkowski spacetime $\man$ and
(\ref{F_epsilon_expansion}) is \emph{not} the $\Vep$ expansion of
the $2$-form $F^\Vep(\VarC^\Vep(Q))$ on $\Bman$.

One way to minimise the complexity of the ensuing calculation is to
adapt a coordinate system on $\UBman$ to $\calJ$. Let
$(\xi^1,\xi^2,\xi^3)$ be a coordinate system on $\UBman$ and let $
\calJ_{123}$ be the corresponding component of $\calJ$: 
\begin{equation}
\label{multi_basic_form_components}
\calJ = \calJ_{123}d\xi^1\wedge d\xi^2\wedge d\xi^3.
\end{equation}
Eliminating $\Brho$ in $J_{[i]}$ in favour of $\calJ$ and $C$ using 
(\ref{multi_Delta}-\ref{multi_proper_charge_density}) and
(\ref{multi_basic_form_components}) yields
\begin{equation*}
J_{[i]}(p) =
\left|\frac{\calJ_{123}(P_{[i]})}{\det({\bf
D}C)(P_{[i]})}\right|\dot{C}(P_{[i]})  
\end{equation*}
where $P_{[i]}=C^{-1}_{[i]}(p)=(\lambda,\xi^1,\xi^2,\xi^3)$ and ${\bf D}C$ is the 
Jacobian of $C^a(\lambda,\xi^1,\xi^2,\xi^3)$.
By definition, $\calJ_{123}$ is independent of $\lambda$ (see
(\ref{multi_basic_form})) so $(\xi^1,\xi^2,\xi^3)$ may be chosen so
that $|\calJ_{123}|=1$.
Hence
\begin{equation*}
\begin{split}
J_{[i]}(p) &=
\frac{1}{|\det({\bf
D}C)(P_{[i]})|}\dot{C}(P_{[i]})\\
&= \frac{1}{|(\det({\bf
D}C)\circ C^{-1}_{[i]})(p)|}(\dot{C}\circ C^{-1}_{[i]})(p)
\end{split}  
\end{equation*}
and so
\begin{equation}
\label{multi_current_final_form}
J^{\Vep a}_{[i]}(p) =
\frac{1}{|(\det({\bf D}\VarC^\Vep)\circ
\VarC^{\Vep -1}_{[i]})(p)|}(\VarCdot{\Vep}\circ \VarC^{\Vep -1}_{[i]})^a(p) 
\end{equation}
where (\ref{multi_VarC_and_C}) and (\ref{multi_change_of_variable})
have been used, ${\bf D}\VarC^\Vep$
is the Jacobian of $\VarC^{\Vep a}(s,\xi^1,\xi^2,\xi^3)$ and ${\bf
  D}C^\Vep = \frac{1}{\Vep}{\bf D}\VarC^\Vep$.

Let $F^{\Vep a}_b = \eta^{ac}F^\Vep_{bc}$ where $F^\Vep =
\frac{1}{2}F^\Vep_{ab}dx^a\wedge dx^b$ and
$\eta^{ac}\eta_{cb}=\delta^a_b$ where
\begin{equation*}
\delta^a_b =
\begin{cases}
1\quad\text{if}\,\,a=b\\
0\quad\text{if}\,\,a\neq b.
\end{cases}
\end{equation*}
The inertial coordinate representations of the Lorentz force equation
(\ref{multi_lorentz_force_epsilon}) for $\VarC^\Vep$ and the
normalisation constraint (\ref{multi_expansion_C_normalisation}) are 
\begin{equation}
\label{multi_lorentz_components}
\VarCddot{\Vep a} = \Vep(F^{\Vep a}_b\circ\VarC^\Vep)\VarCdot{\Vep b},\qquad
\eta_{ab}\VarCdot{\Vep a}\VarCdot{\Vep b} = -\Vep^2
\end{equation}
where $F^\Vep$ is a solution to the Maxwell equations
\begin{equation}
\label{multi_maxwell_epsilon}
dF^\Vep =0,\qquad d\star F^\Vep = -\sum\limits^{N(p)}_{i=0}\star\dual{J^\Vep_{[i]}}
\end{equation}
and $J^\Vep_{[i]}$ is given in (\ref{multi_current_final_form}).

Inserting (\ref{C_epsilon_expansion}) and (\ref{F_epsilon_expansion})
into (\ref{multi_lorentz_components}) and
(\ref{multi_maxwell_epsilon}) and equating equal order terms in $\Vep$
induces a hierarchy of equations for successive approximations to
$(\VarC^\Vep,F^\Vep)$. The first six steps in the hierarchy are:  
\begin{itemize}
\item Adopt an external electromagnetic field $F_{-1}$, i.e. a
solution to the source-free Maxwell equations
\begin{equation*}
dF_{-1} = 0,\qquad d\star F_{-1} = 0.
\end{equation*}
\item Solve
\begin{equation*}
\VarCddotind{a}{0}(Q) = (F_{-1})_b{}^a(p)\, \VarCdotind{b}{0}(Q)
\end{equation*}
for $\VarC_0^a$ subject to
\begin{equation*}
\eta_{ab}\,\VarCdotind{a}{0}(Q)\VarCdotind{b}{0}(Q) = 0
\end{equation*}
where $p=\VarC_0(Q)$ and $(F_{-1})_b{}^a$ is
data obtained in the previous step.
\item The $2$-form $F_0$ is a solution to the Maxwell equations
\begin{equation*}
dF_0 = 0,\qquad d\star F_0 = -\sum\limits^{N(p)}_{i=0}\star\dual{J}_{[i]0}
\end{equation*}
and
\begin{equation}
\label{multi_epsilon_zero_order_current}
J^a_{[i]0}(p) = |\det({\bf
D}\VarC^{-1}_{[i]0})(p)|\,\VarC^{a\prime}_{[i]0}(Q)
\end{equation}
where ${\bf D}\VarC^{-1}_{[i]0}(p)$ is the Jacobian of
$\VarC^{-1}_{[i]0}$ at $p=(x^a)$ and $Q=\VarC^{-1}_{[i]0}(p)$.
\item The first order correction $\VarC_1$ to the map $\VarC_0$ is
obtained from the linear equation
\begin{equation*}
\VarCddotind{a}{1} =
(F_{-1})_b{}^a \VarCdotind{b}{1} +
(F_{-1})_b{}^a{}_{,c} \VarC_1^c \VarCdotind{b}{0}
+ (F_0)_b{}^a \VarCdotind{b}{0}
\end{equation*}
subject to
\begin{equation*}
\eta_{ab}\,\VarCdotind{a}{0}\VarCdotind{b}{1} = 0
\end{equation*}
where maps on $\Bman$ are implicitly evaluated at $Q$ and maps on
$\man$ are evaluated at $p=\VarC_0(Q)$.
Indices following a comma indicate partial differentiation with
respect to the corresponding coordinates so $(F_{-1})_b{}^a{}_{,c} = \frac{\PD}{\PD
x^c}(F_{-1})_b{}^a$. 
\item The $2$-form $F_1$ is a solution to the Maxwell equations
\begin{equation*}
dF_1 = 0,\qquad d\star F_1 = -\sum\limits^{N(p)}_{i=0}\star\dual{J}_{[i]1}
\end{equation*}
and
\begin{equation*}
\begin{split}
J^a_{[i]1} =& |\det({\bf
D}\VarC^{-1}_{[i]0})|
\bigg[-\text{tr}\big({\bf D}\VarC^{-1}_{[i]0}{\bf D}\VarC_1\big)\VarC^\prime_0\\
&\quad
+\big\{\text{tr}\big({\bf D}\VarC^{-1}_{[i]0}{\bf D}^2\VarC_0\big)^T
     {\bf D}\VarC^{-1}_{[i]0}\VarC_1\big\}\VarC^\prime_0\\ 
&\quad
+ \VarC^\prime_1
- {\bf D}\VarC^\prime_0{\bf D}\VarC^{-1}_{[i]0}\VarC_1
\bigg]^a
\end{split}
\end{equation*}
where maps on $\man$ are implicitly evaluated at $p$ and maps on
$\Bman$ are implicitly evaluated at $Q=\VarC^{-1}_{[i]0}(p)$. Inside
the square brackets ${\bf D}\VarC_1$ is the matrix of derivatives of
$\VarC^a_1$ and ${\bf D}\VarC^\prime_0$ is the matrix of derivatives
of $\VarC^{\prime a}_0$ and both should be regarded as linear maps
from the $(\xi^0=\lambda,\xi^1,\xi^2,\xi^3)$ components of vectors
on $\Bman$ to the $(x^a)$ components of vectors on $\man$.
Similarly, inside the square brackets ${\bf D}\VarC^{-1}_{[i]0}$
should be regarded as a linear mapping from the $(x^a)$ components of
vectors on $\man$ to the $(\xi^a)$
components of vectors on $\Bman$. The column vector
$\text{tr}\big({\bf D}\VarC^{-1}_{[i]0}{\bf D}^2\VarC_0\big)$ is
\begin{equation*}
\left(
\begin{array}{c}
\text{tr}\big({\bf D}\VarC^{-1}_{[i]0}{\bf
    D}\PD_\lambda\VarC_0\big)\\
\text{tr}\big({\bf
    D}\VarC^{-1}_{[i]0}{\bf
    D}\PD_{\xi^1}\VarC_0\big)\\
\text{tr}\big({\bf
    D}\VarC^{-1}_{[i]0}{\bf
    D}\PD_{\xi^2}\VarC_0\big)\\
\text{tr}\big({\bf
    D}\VarC^{-1}_{[i]0}{\bf
    D}\PD_{\xi^3}\VarC_0\big)
\end{array}
\right)
\end{equation*}
where ${\bf D}\PD_\zeta\VarC_0$ is the matrix of partial derivatives
of $\PD_\zeta\VarC_0$ where $\zeta=\lambda,\xi^1,\xi^2,\xi^3$.
\item The second order correction $\VarC_{2}$ to $\VarC$ is a solution
  to the equation
\begin{equation*}
\begin{split}
\VarCddotind{a}{2}(Q) &= 
(F_{-1})_b{}^a\VarCdotind{b}{2}
+ (F_{-1})_b{}^a{}_{,c}\VarC^c_1\VarCdotind{b}{1}\\
&\quad +
\frac{1}{2}(F_{-1})_b{}^a{}_{,cd}\VarC^c_1\VarC^d_1\VarCdotind{b}{0}
+ (F_{-1})_b{}^a{}_{,c}\VarC^c_2\VarCdotind{b}{0}\\
&\quad + (F_0)_b{}^a\VarCdotind{b}{1}
+ (F_0)_b{}^a{}_{,c}\VarC^c_1\VarCdotind{b}{0}
+ (F_1)_b{}^a\VarCdotind{b}{0}
\end{split}
\end{equation*}
subject to
\begin{equation*}
2\eta_{ab}\,\VarCdotind{a}{0}\VarCdotind{b}{2}+\eta_{ab}\,\VarCdotind{a}{1}\VarCdotind{b}{1}
= -1
\end{equation*}
where $p=\VarC_0(Q)$.
\end{itemize}
Note that $\VarC^\prime_0$ and $\VarC^\prime_0+\Vep\VarC^\prime_1$ are 
lightlike and $\VarC^\prime_0+\Vep\VarC^\prime_1+\Vep^2\VarC^\prime_2$
is the leading order timelike approximation to $\VarC^\Vep$.

In single-current regions the reduced proper charge density is
$\sqrt{-g(J_{[1]}^\Vep,J_{[1]}^\Vep)}$ on spacetime. Using
(\ref{C_epsilon_expansion}), (\ref{multi_current_final_form}) and
the normalisation condition (\ref{multi_lorentz_components}) on
$\VarC^{\Vep\prime}$ it follows that
$\sqrt{-g(J_{[1]}^\Vep,J_{[1]}^\Vep)}$ converges to $0$ as $\Vep$
tends to $0$. The reduced proper charge density diverges on
interfaces between single-current and multi-current regions and the
above approximation scheme is valid arbitrarily close to such interfaces.
\section{Conclusion}
Continuum models of charged particle beams include the interaction of
matter with its own electromagnetic field and avoid
peculiar phenomena evident in point-particle descriptions of
self-interacting charge, such as self-acceleration and
pre-acceleration. A field-theoretic realisation of a collection of
classical electrons is the ``cold'' charged continuum. It was shown
that the velocity field of the continuum may possess crossing
trajectories and in this case the Eulerian theory is
inconsistent. Such behaviour is not uncommon in physics; compression
waves in inviscid fluids also develop crossing trajectories. However,
inviscid fluids are an idealisation; all normal fluids are viscous to some
extent and ameliorate the problem by forming a shock. On the other
hand, a beam of electrons is very different from a normal fluid and it
was argued that there is no physical reason why the trajectories
modelling an electron beam cannot cross. A Lagrangian theory permitting crossing
trajectories was presented and its ``wall of charge'' solutions were examined and
compared with solutions to the original Eulerian field system.
Finally, an approximation scheme was developed to
analyse ultra-relativistic charge configurations of the Lagrangian
system.

The Lagrangian theory discussed here features an $N$-component
electric current where $N$ is dynamically determined and has a
point-wise dependence on spacetime. It is possible that configurations
with $N=1$ initially may evolve into highly complicated ``turbulent''
configurations where $N$ is arbitrarily large. Further work in this
context may be found in~\cite{GIFT:2006}.
 
The theory developed here has immediate application to high-energy accelerator
physics where ultra-relativistic motion is ubiquitous. The general
approach is also valid in systems where electromagnetic interactions
dominate over collisional processes, such as in
laser-driven plasma wakefield accelerators where the transition between
single and multiple component electron currents (``wave
breaking'')~\cite{nature:2004} has recently received much attention.
\\    
\begin{acknowledgments}
The authors are grateful to the Cockcroft Institute for making this
work possible, to EPSRC for a Portfolio award and to the EC for a
Framework 6 (FP6-2003-NEST-A) award. RWT is also grateful to EPSRC for
a Springboard Fellowship.  
\end{acknowledgments}


\begin{thebibliography}{99}
\bibitem{dirac:1938}
P.A.M. Dirac, Proc. R. Soc. Lond. {\bf A} 167 148--168 (1938).
\bibitem{landau_lifshitz:1962}
L.D. Landau, E.M. Lifshitz, ``The Classical Theory of Fields'',
Pergamon, Oxford (1962).
\bibitem{burton_gratus_tucker:2006}
D.A. Burton, J. Gratus, R.W. Tucker. Ann. Phys. (in
press, available on-line at http://www.elsevier.com).\\
D.A. Burton, J. Gratus, R.W. Tucker. Proceedings of EPAC
2006. 
\bibitem{benn_tucker:1987}
For an introduction to exterior calculus and its application to
electromagnetism see, for example, I.M. Benn, R.W. Tucker, ``An Introduction to
Spinors and Geometry with Applications in Physics'', Adam Hilger (1987).
\bibitem{davidson:2001}
R.C. Davidson, H. Qin, ``Physics of intense charged particle
beams in high energy accelerators'', World Scientific (2001).
\bibitem{li_wohlbier_etal:2004}
X. Li, J.G. W\"ohlbier, S. Jin, J.H. Booske, Phys. Rev. {\bf E} 70,
016502 1--12 (2004).
\bibitem{GIFT:2006}
J. Gratus, D.A. Burton, R.W. Tucker. Proceedings of Conference on
Global Integrability and Field Theory, Cockcroft Institute (2006) (to appear). 
\bibitem{nature:2004}
S.P.D. Mangles, et al, Nature 431 (2004) 535--8,\\
C.G.R. Geddes, et al, Nature 431 (2004) 538--41,\\
J. Faure J, et al, Nature 431 (2004) 541--4.
\end{thebibliography}
\end{document}